\documentclass{article}
\usepackage[utf8]{inputenc}
\usepackage{graphicx}
\usepackage{wrapfig}
\usepackage{physics}
\usepackage{caption}
\usepackage{subcaption}
\usepackage{setspace}
\usepackage{authblk}
\usepackage[hyphens]{url}
\usepackage{hyperref}
\usepackage[nottoc]{tocbibind}
\usepackage{float}
\usepackage{mathtools}
\usepackage{subcaption}
\usepackage{color}
\usepackage[dvipsnames]{xcolor}
\usepackage{breqn}
\usepackage{url}
\usepackage{verbatim}
\usepackage{amsmath}
\usepackage{multicol}
\usepackage[superscript,biblabel]{cite}
\usepackage[margin=0.9in]{geometry}
\usepackage[labelfont=bf,labelsep=space,font=footnotesize]{caption}
\usepackage{siunitx}
\usepackage{lipsum}

\author[1,*]{Antonios M. Alvertis}
\author[1]{Raj Pandya}
\author[2,3]{Claudio Quarti}
\author[4]{Laurent Legrand}
\author[4]{Thierry Barisien}
\author[5]{Bartomeu Monserrat}
\author[6]{Andrew J. Musser}
\author[1]{Akshay Rao}
\author[4]{Alex W. Chin}
\author[2,$\dagger$]{David Beljonne}

\affil[1]{Cavendish Laboratory, University of Cambridge, J.\,J.\,Thomson Avenue, Cambridge CB3 0HE, United Kingdom}
\affil[2]{Laboratory for Chemistry of Novel Materials, University of Mons, Place du Parc, 20, B-7000 Mons, Belgium.}
\affil[3]{Université de Rennes, ENSCR, INSA Rennes, CNRS, ISCR (Institut des Sciences Chimiques de Rennes) - UMR 6226, F-35000 Rennes, France}
\affil[4]{Sorbonne Université, CNRS-UMR 7588, Institut des NanoSciences de Paris, INSP, 4 place Jussieu, F-75005, Paris, France.}
\affil[5]{Department of Materials Science and Metallurgy,University of Cambridge, 27 Charles Babbage Road, Cambridge CB3 0FS, U.K.}
\affil[6]{Department of Chemistry and Chemical Biology, Cornell University, Ithaca, NY 14853}
\affil[*]{e-mail: ama80@cam.ac.uk}
\affil[$\dagger$]{e-mail: David.BELJONNE@umons.ac.be}

\begin{document}

\title{First principles modeling of exciton-polaritons in polydiacetylene chains}

\date{\today}

\maketitle

\section*{Abstract}
    Exciton-polaritons in organic materials are hybrid states that result from the
    strong interaction of photons and the bound excitons that these materials
    host. Organic polaritons hold great interest for optoelectronic
    applications, however
    progress towards this end has been impeded by the lack of a first principles 
    approach that quantifies light-matter interactions in these systems, and which would
    allow the formulation of molecular design rules.
    Here we develop such a first principles approach,
    quantifying light-matter interactions. We exemplify our approach by studying variants of the
    conjugated polymer polydiacetylene, and we show that a large polymer conjugation length is
    critical towards strong exciton-photon coupling, hence underlying the importance of 
    pure structures without static disorder.
    By comparing to our experimental 
    reflectivity measurements, we show that the coupling of excitons to vibrations, manifested by phonon side bands in the absorption, has a strong
    impact on the magnitude of light-matter coupling over a range of frequencies. 
    Our approach opens the way towards a deeper
    understanding of polaritons in organic materials, and we highlight that a 
    quantitatively accurate calculation of the exciton-photon interaction would require accounting
    for all sources of disorder self-consistently.

\section{Introduction}

Excitons are bound electron-hole pairs that appear in semiconductors, and are critical to
the optoelectronic response of these materials. 
These quasiparticles are connected to a finite oscillating polarization that emits an 
electromagnetic wave, which can act on an incident wave, leading to a light-matter
interplay. Once a photon is close to resonance with an exciton, light-matter coupling
can become strong, and the two cannot be treated independently, as it was first realized 
in the pioneering works of Pekar \cite{Pekar1958,Pekar1960} and Hopfield 
\cite{Hopfield1958} more than six decades ago. This coupling between
polarization and photons is captured by introducing the quanta of a new quasiparticle referred to as a
polariton, which describes the propagation of electromagnetic energy in matter\cite{Kalt2019}. 

\begin{figure}[tb]
\centering
\includegraphics[width=0.4\linewidth]{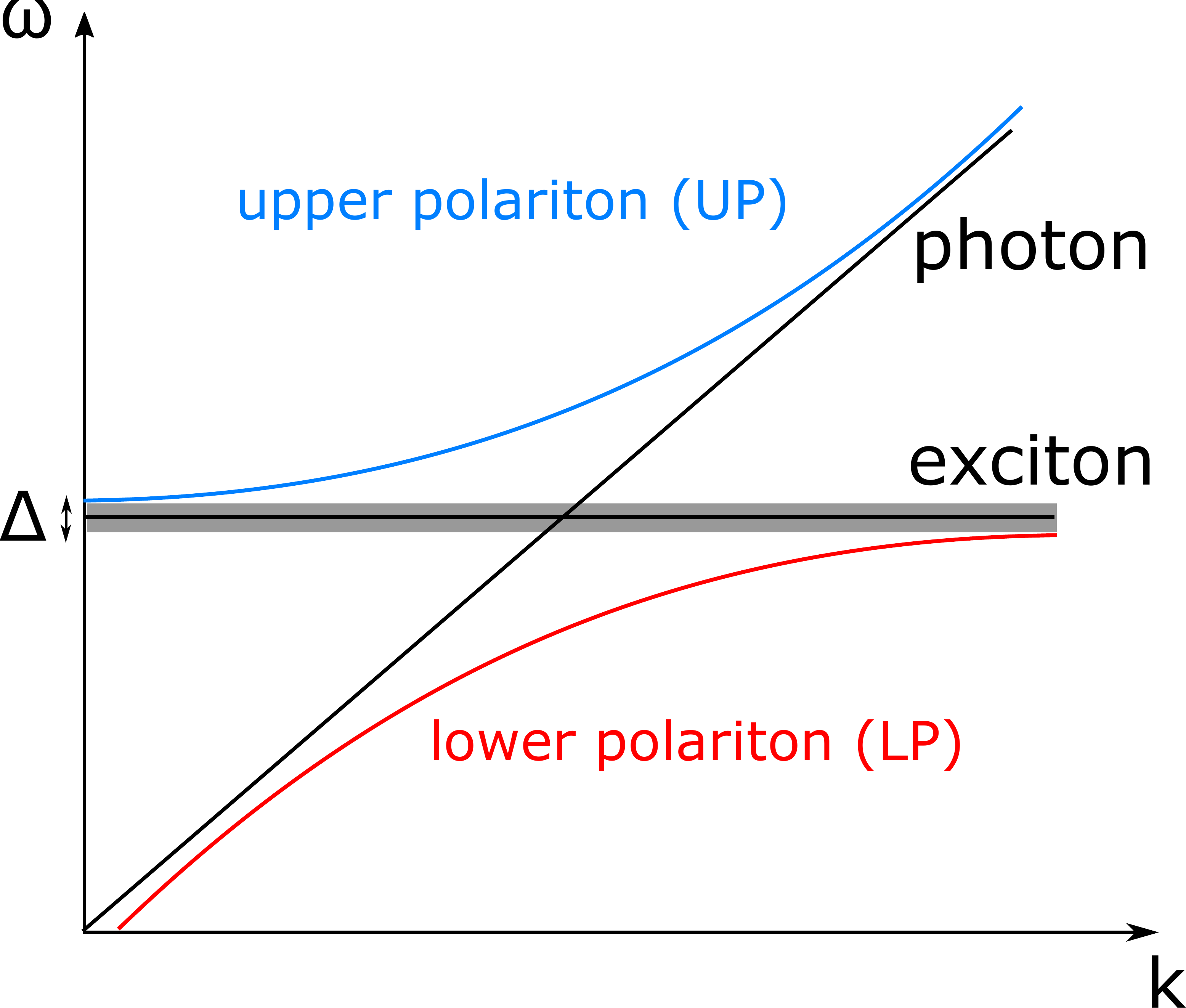}
\caption{The dispersion of a photon in a material intersects with that
of an exciton, and their interaction leads to two new hybrid polariton 
states. The upper and lower polaritons are separated by an energy gap of
width $\Delta$ (shaded region), which is referred to as the stop-band.}
\label{fig:Rabi_schematic}
\end{figure}

Organic semiconductors are a class of materials the light-matter interactions of which are
relevant to a number of promising optoelectronic applications, making the understanding of their
polariton properties interesting from a technological point of view. Their applications include
light-emitting diodes (LEDs) \cite{Reineke2009}, solar cells \cite{Meng2018} and room temperature polariton lasers \cite{Strashko2018,Kena-Cohen2010}. 
Excitons in organic semiconductors are typically strongly bound \cite{Knupfer2003,Kohler2009}
Frenkel-like excitons\cite{Beljonne2013,Berkelbach2013,Yost2014,Coto2015}, and their dispersion
can be considered to be approximately flat\cite{Litinskaya2006} compared to the strong dispersion
of light, as visualized in
Figure\,\ref{fig:Rabi_schematic}. For an incident photon
close to resonance with an exciton state of an organic semiconductor, strong coupling leads to
two new polariton states that are referred to as upper polariton (UP) and 
lower polariton (LP) respectively.
As depicted in
Figure\,\ref{fig:Rabi_schematic}, this results in a stop-band (shaded region),
where no light propagation takes place within the material. The width $\Delta$ of
the stop-band depends
on the magnitude of the exciton-photon interaction. The stop-band is also related to, but is 
distinct from, the so-called Rabi splitting that is commonly used to quantify light-matter 
interactions, and
which is defined as the maximal vertical distance between the upper and lower polariton 
branches\cite{Lidzey1998}. Polaritons in organic semiconductors
have often been studied for molecular systems placed between highly reflecting mirrors
that confine incident light\cite{Polak2020,Kena-Cohen2010,Agranovich2003}. This microcavity effect
leads to a standing electromagnetic wave that interacts with the material system. However, here
we focus on the case of unconfined photons, and not on such microcavity polaritons.

The building block of organic semiconductors is often organic molecules, the large size 
and weak intermolecular van der Waals interactions of which lead to phonon modes with low
frequencies, hence strongly activated at room temperature. These
slow nuclear motions lead to large variations of the electronic
interactions between neighboring molecules \cite{Troisi2005,Troisi2006}. Since excitons 
can spatially extend over several molecules \cite{Cudazzo2015,Refaely-Abramson2017},
such 
low-energy phonons can have a strong influence on their properties \cite{Alvertis2020}.
Similarly, organic 
semiconductors based on conjugated polymers exhibit very low-frequency phonon modes that
are strongly coupled to their excitons \cite{Mahrt1991}. 
However, it is not only vibrations of low energies that couple strongly to excitons,
as in organic systems high-frequency motions such as carbon-carbon stretches have often
been discussed to dominate this interaction\cite{Spano2010,Yamagata2011}. This strong
coupling of excitons to the vibrations of organic semiconductors is expected to also
have an impact on exciton-photon coupling, and indeed it has been shown 
that
vibrations have a strong effect on the dynamics of polariton states
\cite{Somaschi2011,Wu2016,Coles2011,DelPino2018,DelPino2018_2}.

Therefore, understanding the properties of polaritons requires accounting not only for
exciton-photon, but also exciton-vibration interactions, through a Hamiltonian operator
of the form:

\begin{equation}
    H = H_{\text{exc-ph}}+H_{\text{exc-vib}},
\end{equation}

\noindent
where the operator $H_{\text{exc-vib}}$ describes the exciton-vibration interactions.
For a single exciton state with energy $\omega_{\text{exc}}$, 
coupled to a photon of energy $\omega_{\text{ph}}$, the exciton-photon Hamiltonian can be written within the so-called resonance approximation\cite{Haug2004}:

\begin{equation}
\label{eq:exc-ph}
    H_{\text{exc-ph}} = \omega_{\text{ph}}\hat{a}^{\dagger}\hat{a}+\omega_{\text{exc}}\hat{c}^{\dagger}\hat{c}-ig(\hat{a}\hat{c}^{\dagger}-\hat{a}^{\dagger}c).
\end{equation}

\noindent
Here $a^{\dagger}$ ($a$) is the creation (annihilation) operator of the
photon state, and $c^{\dagger}$ ($c$) the creation (annihilation) operator of the exciton. The constant $g$ quantifies the magnitude of the exciton-photon interaction and is related to the width of the stop-band $\Delta$ of Figure\,\ref{fig:Rabi_schematic} via the relationship: $g=\frac{1}{4}\sqrt{ \omega_{\text{ph}}\Delta}$.  

While first principles approaches have been used in several cases to parametrize
the exciton-vibration interactions encoded in $H_{\text{exc-vib}}$ \cite{Schroder2019,Alvertis2019}, the same cannot be said for the exciton-photon Hamiltonian of equation\,\ref{eq:exc-ph}. To the best of our knowledge, a comprehensive
\emph{ab initio} method for modeling the stop-band of
organic systems is thus far missing. Naturally, experiment can inform
computational models on the range of values of $\Delta$, however this
does not allow for a truly predictive approach of modeling light-matter
interactions in these systems. It also remains unclear how vibrational 
effects affect the stop-band. 

Here we develop a theoretical framework for simulating
exciton-photon interactions in organic systems fully from 
first principles.  We exploit the relationship of the microscopic relation 
equation\,\ref{eq:exc-ph} to the classical macroscopic theory of polaritons to 
provide a first principles extraction of the stop-band width $\Delta$.  
We exemplify our approach by studying variants of the conjugated polymer 
polydiacetylene (PDA) (Figure\,\ref{fig:PDA_schematic}), and we show that 
the conjugation length of the polymer chains has a large impact on the magnitude of 
exciton-photon interactions. Exciton properties of PDA are obtained from first principles
by using accurate Green's function based methods ($GW$-BSE) \cite{Lars1965,Rohlfing1998,Rohlfing2000}, which are known to be appropriate for
the study of conjugated polymers\cite{Rohlfing1999}, due to successfully capturing
the large intrinsic correlation length of the electrons and holes that participate
in exciton states. 
We compare the simulated
results to
experimental reflectivity measurements of the stop-band width, and we find the two to be
in reasonable agreement. However, by only considering the effect of the main exciton peak on incident photons, the internal structure of the stop-band as it
appears in reflectivity measurements is not reproduced in the calculation. By coupling $GW$-BSE
calculations for the excitons, to non-perturbative finite differences methods for the vibrations 
\cite{Kunc1982,Monserrat2018}, we show that phonon effects have an important effect on the  internal stop-band structure.

The structure of this paper is as follows. Section\,\ref{theory} provides
an overview of our theoretical approach for describing light-matter 
interactions. 
We apply our methodology to disorder-free PDA in 
section\,\ref{disorder-free}.  An overview of the studied systems
is also provided in this section, alongside a discussion of their exciton 
properties and the experimental determination of the magnitude of light-matter interactions. We then
proceed to discuss the role of static disorder in the sense of limiting
the infinite conjugation length of disorder-free chains, in 
section\,\ref{static-disoder}. The impact of dynamic disorder (vibrations) on 
light-matter interactions is discussed in section\,\ref{dynamic-disorder},
along with a detailed outline of the employed methodology for accounting
for its effect. 
We summarize the entirety of our results and discuss potential future avenues in
the Conclusions of section\,\ref{conclusions}.

\section{Theoretical background}
\label{theory}
Consider an electromagnetic wave $\mathbf{E},\mathbf{B}$ interacting 
with a material of dielectric $\epsilon(\omega)$. The wave equation for the 
electric field in frequency space reads:

\begin{equation}
\label{eq:wave_eq}
    \nabla^2 \mathbf{E}(\mathbf{r},\omega)+\frac{\omega^2}{c^2}\epsilon(\omega)\mathbf{E}(\mathbf{r},\omega)=0. 
\end{equation}

\noindent
For the moment, and for the sake of simplicity, let us consider a bulk isotropic material, for which the electric field can be treated as a plane wave:

\begin{equation}
\label{eq:electric_field}
    \mathbf{E}(\mathbf{r},\omega)= E \cdot e^{i(\mathbf{k}\cdot \mathbf{r}-\omega t)}
\end{equation}

\noindent
By substituting equation~\ref{eq:electric_field} into~\ref{eq:wave_eq} we obtain:

\begin{equation}
    [-k^2+\frac{\omega^2}{c^2}\epsilon(\omega)]E \cdot e^{i(\mathbf{k}\cdot \mathbf{r}-\omega t)}=0.
\end{equation}

\noindent
The transverse eigenmodes of the system satisfy the condition $E \neq 0$, therefore:

\begin{equation}
\label{eq:polariton_equation}
    c^2k^2=\omega^2\epsilon(\omega)
\end{equation}

\noindent
It thus becomes obvious that if one obtains the dielectric response $\epsilon(\omega)$ of a material, the dispersion of the light field in the
material, i.e. 
the
polariton dispersion, can be obtained by solving equation~\ref{eq:polariton_equation}. This method is 
agnostic to the level at which the electronic structure is computed, and our $GW$-BSE methodology provides but 
one possibility. 

Let us now consider the so-called resonance approximation to equation~\ref{eq:polariton_equation},
as a useful way of gaining physical intuition on the origins of the stop-band. However,
we emphasize that we do not make use of this approximation to obtain the simulated results
presented in this paper, and it only serves as a way of developing intuition. 
Unless we are at resonance with an excited state $\omega=\omega_o$, the imaginary part of the dielectric
response is negligible, therefore equation~\ref{eq:polariton_equation} can be approximated as:

\begin{equation}
    \label{eq:resonance_approximation}
    c^2k^2=\omega^2\text{Re}(\epsilon(\omega)).
\end{equation}

\noindent
Equation~\ref{eq:resonance_approximation} shows
that there will only exist real solutions for the wavevector $k$ once $\text{Re}(\epsilon(\omega))$ is positive. While this is generally the case,
$\text{Re}(\epsilon(\omega))$ can become negative in the vicinity of an optical
transition $i$ with a high oscillator strength $f_i$. This can be seen 
within a Lorentz oscillator model of the dielectric response \cite{JacksonJohnDavid1975Ce/J}:

\begin{equation}
\label{eq:Lorentz_model}
\epsilon(\omega)=\epsilon_0 +\frac{Ne^2}{m}\sum_i \frac{f_i}{\omega_i^2-\omega^2-i\omega\gamma_i}.
\end{equation}

\noindent
The real part can be written as:

\begin{equation}
    \text{Re}(\epsilon(\omega))=\epsilon_0 +\frac{Ne^2}{m}\sum_if_i\frac{\omega_i^2-\omega^2}{(\omega_i^2-\omega^2)^2+\omega^2\gamma_i^2}
\end{equation}

\noindent
where $\epsilon_0$ is the background dielectric response, $N$ the number of molecules per unit volume, $m$ the electron mass and $\gamma_i$
the dephasing rate. In the vicinity of a transition $i$, $\omega=\omega_i+\delta \omega$, and only keeping terms in first order of $\delta \omega$ we obtain the condition for $\text{Re}(\epsilon(\omega))<0$
to be:

\begin{equation}
\label{eq:condition}
    f_i>(1+\frac{\omega_i}{2\delta \omega})\cdot \frac{\gamma_i^2}{\omega_p^2},
\end{equation}

\noindent
where $\omega_p^2= \frac{Ne^2}{\epsilon_0m}$ the plasma frequency of a 
material. The width of the frequency region over which 
condition\,\ref{eq:condition} holds and thus 
$\text{Re}(\epsilon(\omega))<0$, defines the stop-band width $\Delta$.

Therefore, the brighter an optical transition, the more likely it is that it
will lead to a forbidden region for light propagation in the material. Since
the oscillator strength of a material satisfies the sum rule:

\begin{equation}
\label{eq:sum_rule}
    \sum_i f_i = Z,
\end{equation}

\noindent
with $Z$ the number of electrons per molecule, we conclude that materials
with fewer bright optical transitions are more likely to satisfy
condition\,\ref{eq:condition} and
lead to strong
light-matter interactions. This provides the motivation for studying extended one-dimensional systems like PDA,
as elaborated in the following section.

Before proceeding to the results section, we would like to highlight two points regarding the range of applicability of the
presented methodology. First of all, this has been discussed in the context of a general
Lorentz oscillator, and the frequencies $\omega_i$ appearing in equation\,\ref{eq:Lorentz_model}
need not correspond to exciton states. While here we focus on exciton-polaritons, optical
phonons and plasmons can also form hybrid states with light \cite{Dai2014,Zayats2005}, and the
presented theoretical framework could be used to describe such phenomena. Secondly,
the polariton properties we calculate correspond to the bulk of an infinite material
structure. However, these states can also form at the surface of a material,
only propagating along the interface between different media. These states are referred to as
surface polaritons, and have a finite dispersion within the stop-band of the infinite `bulk' polariton states that we study here \cite{Agranovich1982}. Our theoretical framework does not
capture such surface effects, but could be extended in that direction in the future.

\section{Application to disorder-free PDA}
\label{disorder-free}

\begin{figure}[tb]
\centering
\includegraphics[width=0.4\linewidth]{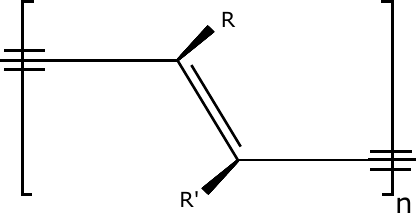}
\caption{The structure of a diacetylene monomer, which generally coincides with the unit cell of an infinite and periodic crystal of polydiacetylene.}
\label{fig:PDA_schematic}
\end{figure}

\begin{figure*}[tb]
\centering
\includegraphics[width=0.9\linewidth]{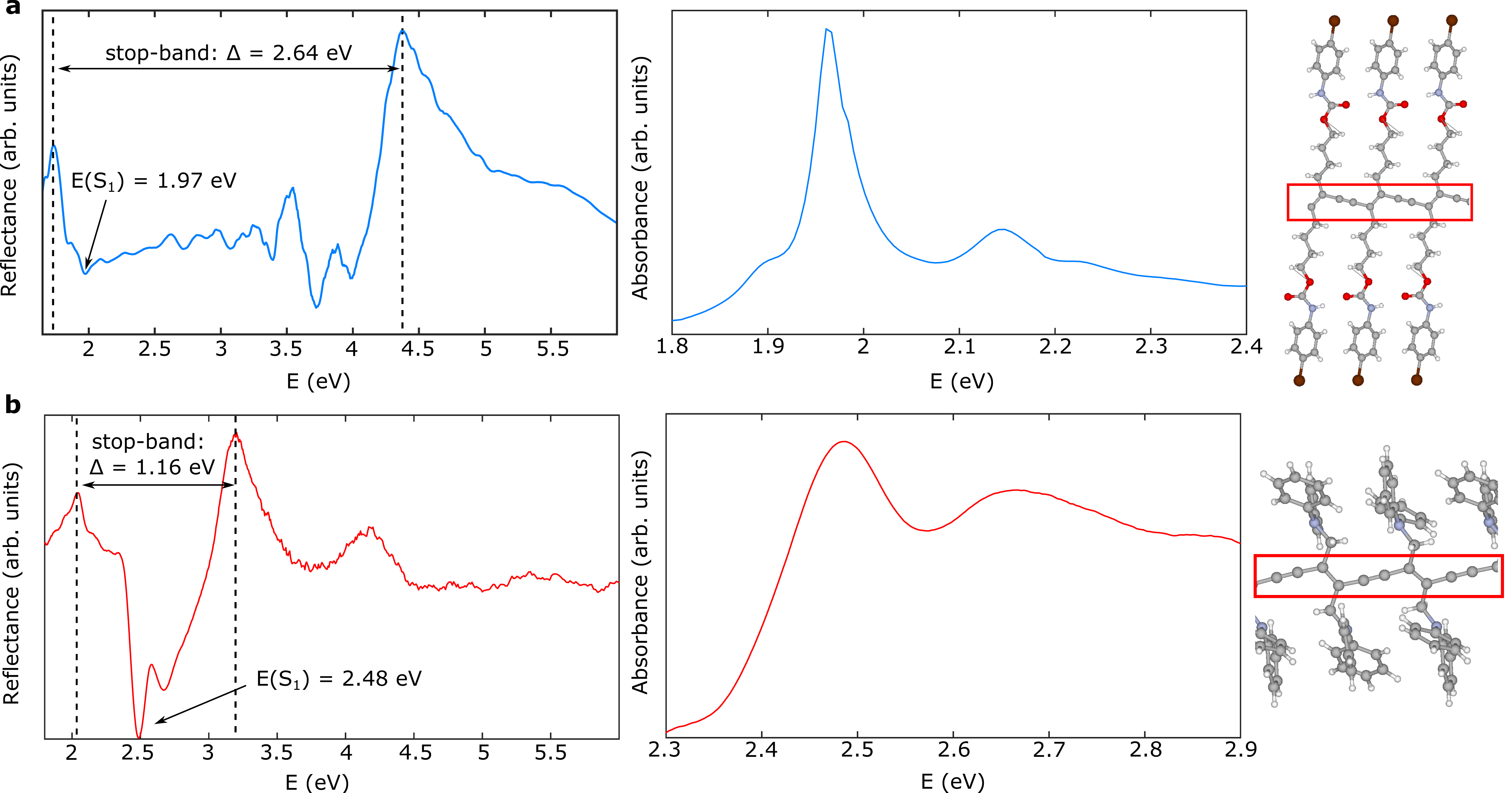}
\caption{Experimental reflection and absorption spectra of two studied variants of polydiacetylene:
\textbf{a} blue chains and \textbf{b} red chains. The stop-band width of the two structures is
approximately determined from the reflectivity spectra.
The polymer structures are given alongside
the experimental plots, with the polymer backbone highlighted in red rectangles, in order to 
distinguish it from the side-chains. Carbon atoms are denoted in light gray, hydrogen in white, oxygen in red,  nitrogen in light blue and bromine in brown.  
}
\label{fig:PDA_structures}
\end{figure*}

Polydiacetylene (PDA) is a conjugated polymer with an alternating sequence 
of
single-double-single-triple carbon-carbon bonds, as visualized in 
Figure\,\ref{fig:PDA_schematic}. These materials are known to give
very high-quality crystals, with an almost perfect alignment of the
polymer chains parallel to each other and free from conformational defects, due to the topochemical reaction they originate from \cite{SCHOTT19873}. Variants of PDA are traditionally named after the color of the
transmitted light through a film of the material. Here we study two PDA variants shown in Figure\,\ref{fig:PDA_structures}, alongside their absorption and reflection spectra.
We refer to the
structure of Figure\,\ref{fig:PDA_structures}a that absorbs at lower energies as 
`blue' PDA \cite{Spinat1985} and the one of Figure\,\ref{fig:PDA_structures}b  that absorbs 
at higher energies as `red' \cite{Enkelmann1980,Barisien2007}. It is important to 
note that the two types of PDA chains and their optical responses should in no way be
associated to varying levels of disorder in their structures but are largely 
determined by their backbone geometries (see below)\cite{Schott2006}. 

\begin{figure}[tb]
\centering
\includegraphics[width=0.4\linewidth]{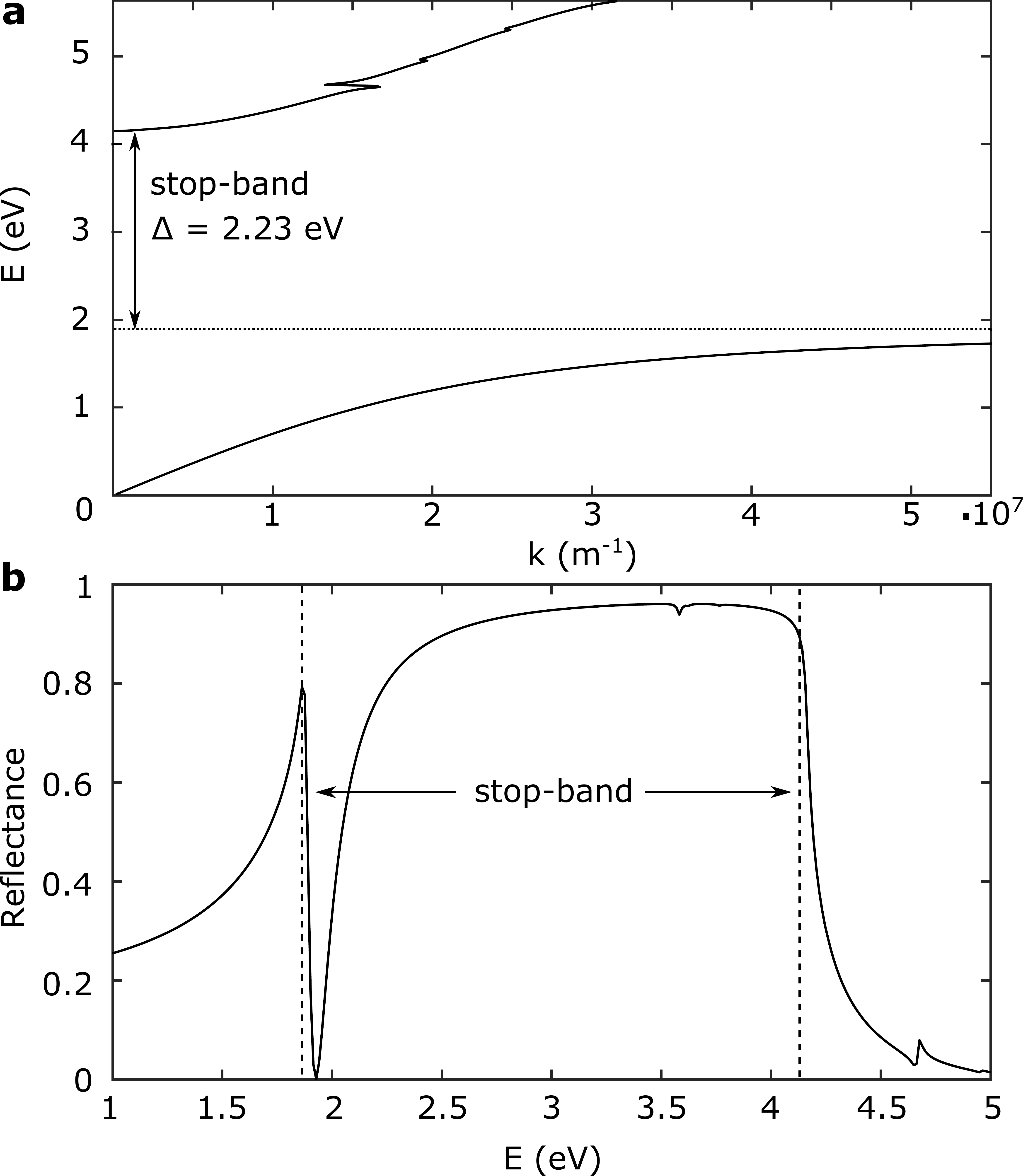}
\caption{Calculated \textbf{a} polariton dispersion and \textbf{b} reflectance of the model blue PDA chains, highlighting the stop-band of width $\Delta=2.23$\,eV.}
\label{fig:polaritons_static}
\end{figure}

We determine
the stop-band of the two PDA variants by measuring their reflectivity
spectra presented in Figure\,\ref{fig:PDA_structures}, for near-normal incidence of light. Details of the experimental 
measurements are given in Appendix A. If we were to disregard the absorption
from the exciton over a finite region of frequencies, then every incident photon
with an energy within the stop-band region would be reflected, giving a reflectance of $R=1$\cite{Philpott1980}, since no propagation of light can take place within the forbidden region. However, the finite damping $\gamma$ that appears
in the dielectric function of equation\,\ref{eq:Lorentz_model} due to the absorption of
an exciton, leads to weaker reflectance and a `gap' within the forbidden region of the stop-band\cite{Meskers2016,Kalt2019} (as also seen in Figure\,\ref{fig:polaritons_static}b for a model PDA system). Further gaps to the high reflectivity of the stop-band can appear due to
disorder or additional bright exciton peaks. Therefore, it is not always
the case that one can identify the stop-band as a region of high reflectance, and its
internal structure can be rich. This is the case with the reflectivity spectra of Figure\,\ref{fig:PDA_structures}. Nevertheless, we can still identify the width of the stop-band as
the distance of two peaks in the reflectivity spectra: the one which appears on the low-energy
side of the dip due to the absorption from an exciton state, and the one for which the
reflectance reaches values at least as large as the maximum of the first peak.
This way of determining the stop-band is similar to previous work\cite{Meskers2016}, and strong light-matter
coupling in squarylium \cite{Tristani-Kendra1983} provides a particularly good example of
using the reflectivity measurements to that end.

Our reflectivity measurements show that the stop-band width is $2.64$\,eV and
$1.16$\,eV for the blue and red chains respectively. We discuss the origin of this large
difference between the red and blue chains with the aid of simulations below.
The measured values are within the expected
range, as the stop-band width has been shown to surpass values of $1-2$\,eV in several organic
materials\cite{Meskers2016}. In particular, it has been shown that 
for near-normal incidence of light on a lattice of aligned dipoles, values of this order of
magnitude might be obtained\cite{Meskers2016}. Given the close-to-perfect alignment of the polymer chains in PDA,
as well as the large oscillator strength of their lowest bright exciton state, it is not 
surprising that light-matter coupling is strong. 

Let us now examine some of the details of the experimental reflectivity spectra. 
For the blue chains, the reflectance dips around the energy of the bright
singlet exciton state at $1.97$\,eV, and a vibronic peak is also weakly resolved at $2.15$\,eV.
Furthermore, a small dip appears at
$1.85$\,eV, and this has been discussed to be due to minority chain populations
with different geometries, which are present within the samples\cite{Schott2006}.
Additional structure appears at energies of $2.72$\,eV and $3.06$\,eV. It is
currently not clear what these peaks correspond to, however 
the side-chains that are connected to the conjugated PDA structure can host exciton
states of their own, potentially leading to some of the observed features. 
At higher energies, the reflectance
shows an increase at approximately $3.5$\,eV and then decreases again. This is due to 
exciton states of unpolymerized diacetylene monomers which are known to appear in this energy
range\cite{Schott2006}. Finally, there is a stark increase of the reflectance as photon
energies approach $4.4$\,eV, which is then followed by a continuous drop in its value, behavior
characteristic of the stop-band edge. For the red PDA, a large dip in the reflectance appears
at $2.48$\,eV where the bright exciton state of this structure absorbs,
and a vibronic peak is then present at $2.68$\,eV. 
The reflectance then increases steadily until
reaching the edge of the stop-band. On the low-energy side of the main exciton peak, some 
additional structure is present, and similar to the case of blue PDA, we attribute this
to minority chain populations. 

\begin{figure}[tb]
\centering
\includegraphics[width=0.4\linewidth]{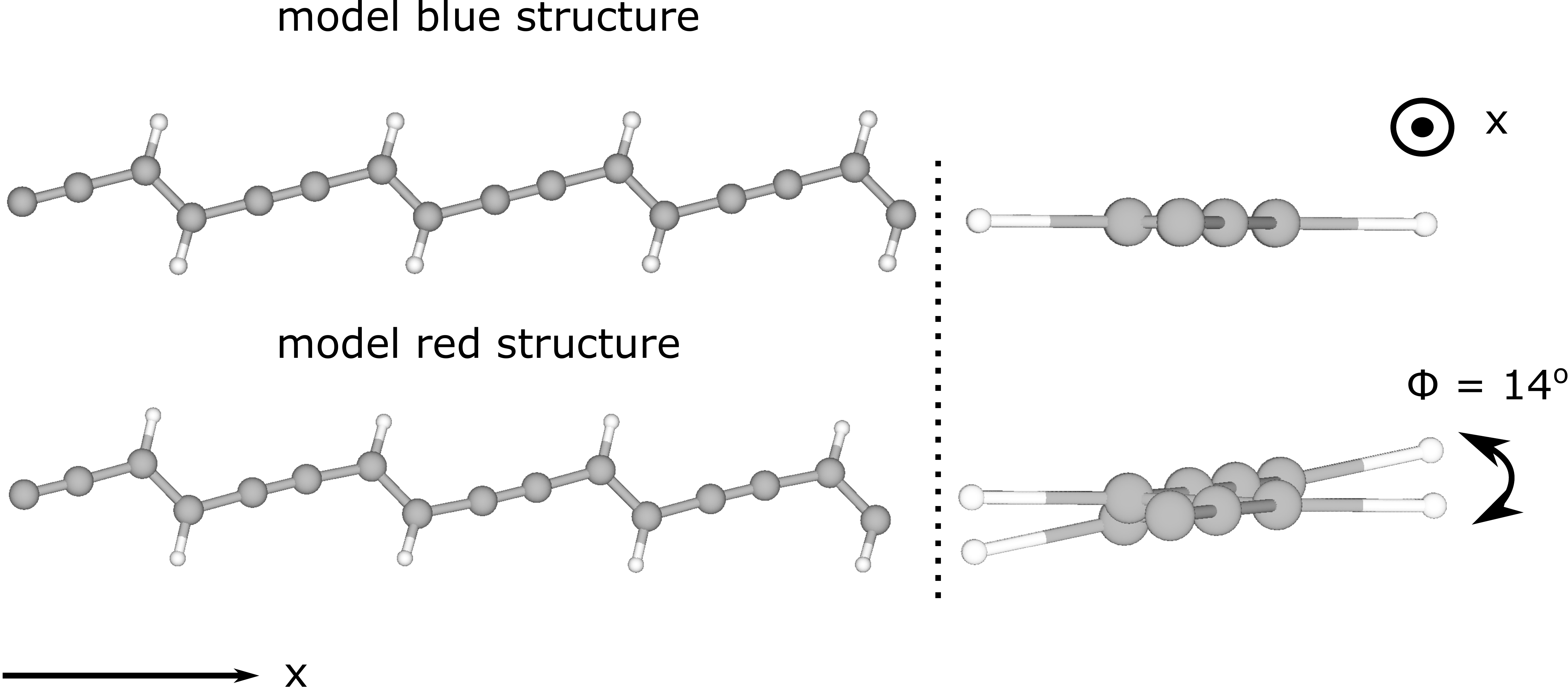}
\caption{Model PDA structures, constructed by substituting the side-chains
of the real materials of Figure\,\ref{fig:PDA_structures} with hydrogen atoms. While the two model systems might look identical when viewed from
the side (left), viewing them in the direction in which the chains lie (right) emphasizes that the blue structure is planar, while a finite
torsion occurs between subsequent monomers of the red structure.}
\label{fig:model_structures}
\end{figure}

The PDA variants of Figure\,\ref{fig:PDA_structures} contain a very large 
number of atoms in their unit cells, due to the size of the side-chains
connected to the conjugated structure (highlighted with a red rectangle in Figure\,\ref{fig:PDA_structures}).
In order to make the computational problem of obtaining accurate exciton
properties for PDA feasible, we resort to studying two
simplified model PDA chains, where the side-chains have been substituted
by hydrogen atoms, as visualized in Figure\,\ref{fig:model_structures}.
The lattice parameters, as well as the interatomic distances within the conjugated
chains for the two model PDAs, are kept to be the same as for the respective
real structures.
When viewing the model PDA chains sideways as on the left of Figure\,\ref{fig:model_structures} these systems look identical. However,
by looking into the direction along which the chains extend, it becomes
clear that the model red structure is not planar, and there exists a finite
torsional angle of $14^{\text{o}}$ between subsequent diacetylene monomers. 
The effect of this torsion has been discussed in detail previously \cite{Filhol2009},
and shown to lead to significant differences in the optoelectronic response
of the two systems. In crystalline PDAs the side groups, through their mutual 
organization, pilot  the chain conformation that in turn governs the electronic 
properties. However, the side groups themselves do not participate to the conjugated
system, justifying the substitution by hydrogen atoms. Nevertheless, this 
simplification leads to a renormalization of exciton energies. This is due to the
the strong dielectric mismatch introduced between the conjugated chain and the 
outside medium, as discussed below.

\begin{figure}[tb]
\centering
\includegraphics[width=0.4\linewidth]{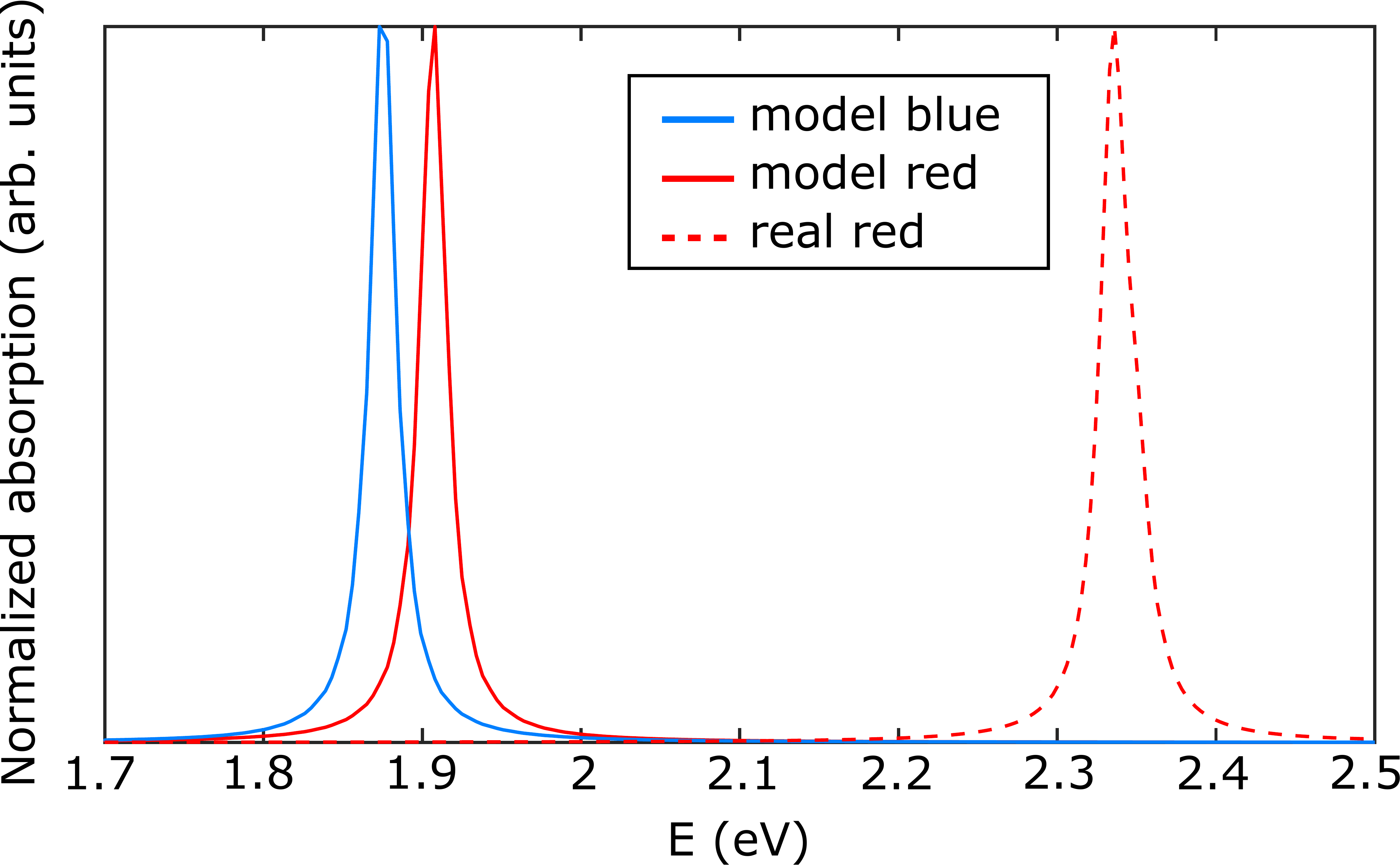}
\caption{Normalized absorption of the two model PDA chains and the real red structure.}
\label{fig:PDA_static_absorption}
\end{figure}

The absorption of the model PDAs and real red system as computed from $GW$-BSE is given in Figure\,\ref{fig:PDA_static_absorption}. The computational details for the electronic structure calculations are given
in Appendix B. We were not able to obtain converged exciton properties for the real
blue PDA system, and we exclude it from any further computational analysis. The first thing to observe from Figure\,\ref{fig:PDA_static_absorption} is that there is only a single
bright transition for all the studied systems, at least in the idealized case of the infinite
periodic material. Therefore, PDA provides
an ideal system for satisfying condition\,\ref{eq:condition} and observing strong light-matter interactions, since
the oscillator strength is not shared between several transitions according
to the sum rule of equation\,\ref{eq:sum_rule}. It should however be noted that the real systems
has some additional exciton states with small but non-vanishing oscillator strength at energies above $3$\,eV, which are associated with the side-chains.

\begin{figure}[tb]
\centering
\includegraphics[width=0.4\linewidth]{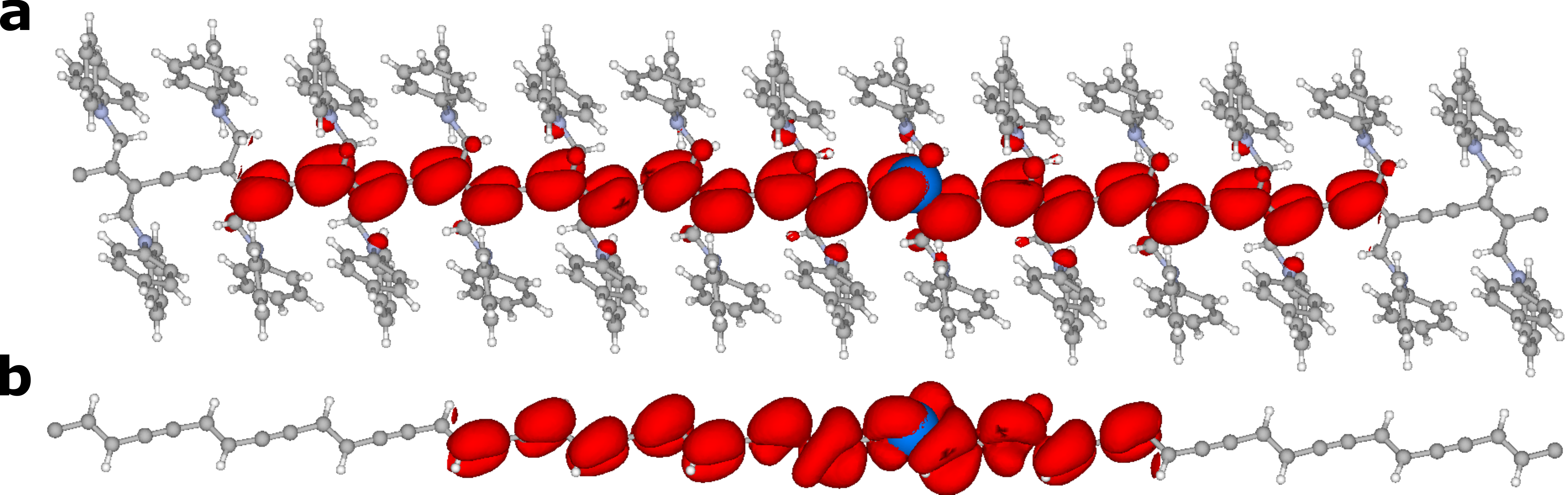}
\caption{Exciton wavefunction of \textbf{a} the real red PDA chains and 
\textbf{b} the model red chains. The electron density is visualized in red
for a hole fixed at the position indicated by a blue sphere.}
\label{fig:exciton_WF}
\end{figure}

Moreover, it is interesting to note the difference in the exciton energy
between the real and model red structures. Since the polymer backbone
is identical in both cases, it is only the side-chains that can be causing
the observed difference. The side-chains of the real red PDA consist
of two phenyl rings, which are highly polarizable and hence provide significant screening to the Coulomb interaction that binds electron-hole
pairs together. In contrast, in the model system the field lines of this Coulomb interaction are not
screened by a medium, since the side-chains have been substituted by hydrogen atoms. Consequently, the exciton is more strongly bound
in the model chain, further stabilizing its energy. For the real blue system, the side-groups
largely consist of non-polarizable alkyl chains, hence the screening they provide is much weaker
compared to the red PDA. This is reflected by the fact that the exciton energy of the model blue
system, where the side-chains have been substituted by hydrogen atoms, closely reproduces the
experimental value from the absorption spectrum of Figure\,\ref{fig:PDA_structures}a.
Table\,\ref{table:binding_energy} summarizes the values for the fundamental
gaps of the chains, along with the energy of the excitons and their binding
energy. The exciton binding energy of the real red system is equal to $0.52$\,eV, while for both model chains it is greater than $2$\,eV, emphasizing the large impact of the screening provided by the polarizable side-chains. This effect can be further emphasized by visualizing the exciton 
wavefunctions of the real and model red chains in 
Figure\,\ref{fig:exciton_WF}. The electron density is visualized in red,
for a hole fixed at the position indicated by a blue sphere. As expected,
the lack of side-chain screening of the Coulomb interaction between the 
electron and hole in the case of the model chain leads to an exciton that
is spatially more confined than in the real system. As far as the relative 
electron-hole motion is concerned, we also reach a good agreement with the results 
from electroabsorption studies that provide an exciton bohr radius of several 
nanometers, corresponding to an exciton covering a few monomer units in the blue and
red conformations \cite{Horvath1996,Barisien2007}. 

Having obtained $\epsilon(\omega)$ from $GW$-BSE electronic structure 
calculations of the three materials, we proceed to
obtain the polariton
dispersion. Due to the almost perfect alignment of PDA chains along one axis ($x$),
we use the polariton dispersion relation\cite{Meskers2016}:

\begin{equation}
    \label{eq:resonance_approximation_anisotropy}
    c^2k^2=\omega^2\epsilon_{xx}(\omega),
\end{equation}

\noindent
Here $\epsilon_{xx}(\omega)$ is the only non-zero component of the dielectric tensor
of the material, and in the $y$ and $z$ directions, the vacuum dispersion holds: 
$\omega^2=k^2$. We obtain the component $\epsilon_{xx}(\omega)$ of the dielectric tensor,
by solving the Bethe-Salpeter equation for incident light normal to the chains, which 
approximately emulates the experimental conditions used to obtain the reflectivity spectra of Figure\,\ref{fig:PDA_structures}.
Figure\,\ref{fig:polaritons_static}a visualizes the computed dispersion
for the example case of the model blue chains, highlighting the resulting 
stop-band of $2.23$\,eV. The $\Delta$ values obtained for all studied systems are 
are summarized
in Table\,\ref{table:Rabi_static}, and the experimental values are given for comparison. It should be emphasized
that even when we include the effect of exciton dispersion on the polariton states, the
resulting theory values for the stop-band width remain largely unchanged. We see from Table\,\ref{table:Rabi_static} that the model blue chains have significantly larger stop-band
width than the model red system. Since the only significant difference between these structures
is the torsional angle between subsequent diacetylene monomers (see Figure\,\ref{fig:model_structures}), we conclude that the magnitude of this torsion has a
strong effect on the magnitude of light-matter interactions. Indeed the finite torsion
of the model red system leads to deviations from the perfect dipole alignment of 
model blue PDA, and which is crucial to facilitating strong coupling to photons. The computed
values for the stop-band width are in reasonable agreement with the experimental values of
Figure\,\ref{fig:PDA_structures}, and the large difference between the blue and red
structures is captured. 

Furthermore,
the comparison of the predicted $\Delta$ values for the real and model red systems suggests that
the introduction of polarizable side-chains leads to weaker exciton-photon interactions. This is likely 
due to two reasons. Firstly, the dielectric screening that they provide to electron-hole pairs
(Figure\,\ref{fig:exciton_WF}) leads to a more weakly bound exciton, which has a higher transition
energy
compared to the model system. Hence the condition of equation \,\ref{eq:condition} becomes more 
difficult to satisfy. This argument only holds for polarizable side-chains that provide 
significant screening, which is not the case for the real blue PDA, justifying the use of
the model blue system to approximate its stop-band in Table\,\ref{table:Rabi_static}.
Moreover, upon the introduction of side-groups, a few exciton states with small but non-vanishing oscillator strength appear
at energies above $3$\,eV, hence reducing the brightness of the main exciton peak according to
the sum rule of equation\,\ref{eq:sum_rule}, and also making the frequency range that satisfies the condition of equation\,\ref{eq:condition} narrower.

The values predicted by the calculations for the stop-band width are in reasonable agreement
with experiment, as summarized in Table\,\ref{table:Rabi_static}. We believe that the differences
are largely due to our modeling of a perfect periodic structure and
ignoring exciton resonances which are present in a realistic film, for
example due to unpolymerized diacetylene monomers and minority chain populations at different
geometries. Each of these weak transitions couples to photons and has a (narrow) stop-band
of its own. The overlap of the different stop-bands due to the various weak exciton resonances
with that of the main exciton peak leads to an overall wider stop-band\cite{Kalt2019}. In agreement with this line of thought, the predicted values underestimate experiment for both
materials. 

For normal incidence
of light, the reflectance can be calculated as:

\begin{equation}
    \label{eq:reflectance}
    R(\omega)=\frac{(n(\omega)-1)^2}{(n(\omega)+1)^2},
\end{equation}

\noindent
where $n(\omega)$ the complex refractive index of the material. For the model blue chains
this is plotted in Figure\,\ref{fig:polaritons_static}b, where the stop-band is clearly resolved.
The absorption from the exciton state results in a low-reflectance gap within the stop-band, which is in agreement
with experiment. However, the computed reflectivity spectrum looks qualitatively very different
from the experimental one of Figure\,\ref{fig:PDA_structures}a. This is due to entirely
ignoring the effects of disorder in the calculation. The presence of minority chain
populations with different geometries, as well as unpolymerized diacetylene monomers are ignored,
and the exciton states associated with these would lead to more gaps in the reflectivity.
For the model chains we ignore excitons
associated with the side-chains, which would also introduce additional structure within the stop-band.
We will see later that dynamic disorder
is another important factor towards determining the internal structure of the stop-band. 

Before moving on to consider the effects of disorder, we would like to comment on
the speed of energy transfer in these materials, as predicted by the
results of this section. First, let us consider the case of purely excitonic
motion. The quantity that we need to compute in order to quantify the
speed of band-like exciton motion is the exciton group velocity:

\begin{equation}
\label{eq:group_velocity}
    v_{\text{g,exc}} = \frac{1}{\hbar}\cdot \frac{\partial E_{\text{exc}}}{\partial q},
\end{equation}

\noindent
where $E_{\text{exc}}$ the exciton energy and $q$ the exciton wavenumber. 
Following photoexcitation, a wavepacket is generated on the exciton 
potential energy surface, the momentum of which is narrowly distributed
around $q=0$, since light induces close-to-vertical transitions within the Franck-Condon approximation. The center of mass of this
$\Gamma$-point exciton will then travel with the velocity of 
equation\,\ref{eq:group_velocity}. We are interested in energy 
transfer in the direction of the chains, which we define to lie along the
x axis, we thus approximate $v_{\text{g,exc}}$ as the average along the
respective dimension of reciprocal space. To this end, we calculate 
the exciton energy at the band-edge $X$, using a supercell calculation that
is outlined in Appendix B, and we obtain:

\begin{equation}
    v_{\text{g,exc}} \approx \frac{1}{\hbar}\cdot \frac{E_{\text{exc}}(X)-E_{\text{exc}}(\Gamma)}{q(X)-q(\Gamma)}=3\cdot 10^5 \hspace{0.1cm}\text{m/s},
\end{equation}

\begin{figure*}[tb]
\centering
\includegraphics[width=\linewidth]{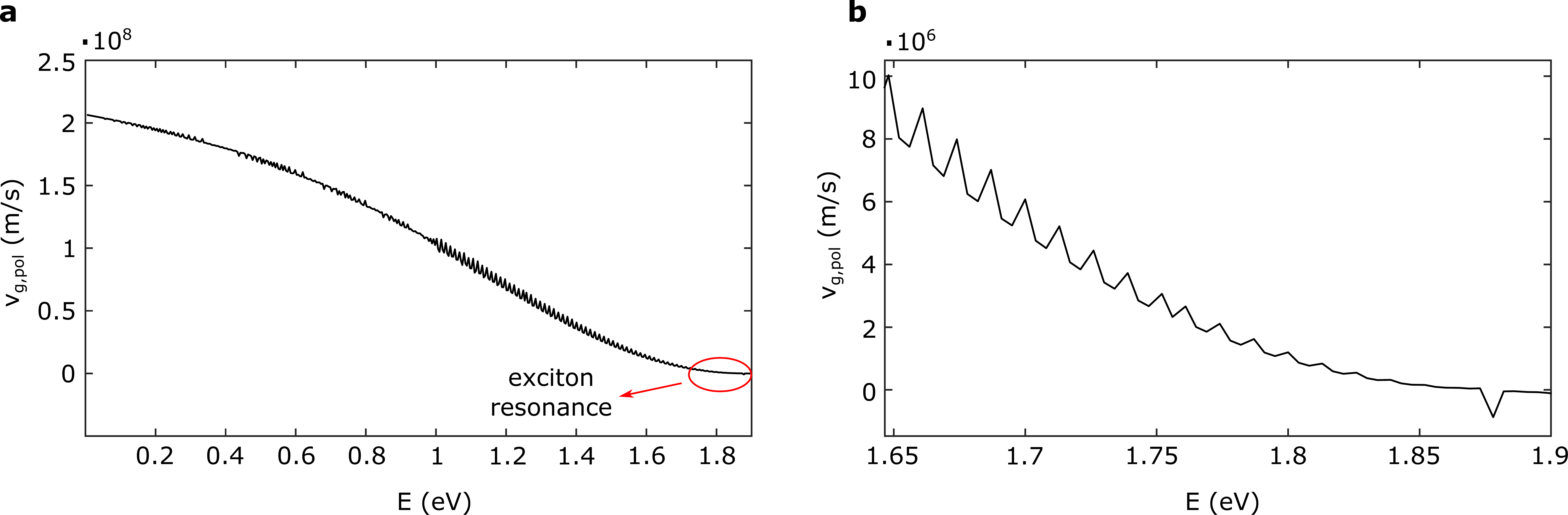}
\caption{Polariton group velocity in model blue chains, for different energies of the incident light. Panel \textbf{a} shows a large region of
frequencies, whereas panel \textbf{b} provides a closer view of the region
close to the resonance with the exciton.}
\label{fig:groupv}
\end{figure*}

Now let us consider the group velocity of a polariton $v_{\text{g,pol}}$ close to the exciton
resonance, which can be computed from the solution of equation\,\ref{eq:resonance_approximation_anisotropy} for different energies of the incident
light. The result for the model blue chains is given in Figure\,\ref{fig:groupv}a; at low frequencies, the speed of light is $c=c_o/n$. However, as we approach the energy of the exciton transition, 
the exciton content of the lower polariton state becomes larger, reducing 
the speed of propagation. Figure\,\ref{fig:groupv}b provides a closer view
to the region around resonance, and we find that the polariton propagates
with a speed of the order of $10^6$\,m/s, which is an order of magnitude
faster than purely excitonic band-like propagation. One might observe that very close to the exciton resonance the group velocity obtains negative values, which
is unphysical. This is due to the fact that in this narrow region,
the material exhibits anomalous dispersion and the group velocity is 
generally not a useful concept in these cases \cite{JacksonJohnDavid1975Ce/J}.

\begin{table}[t]
\centering
  \setlength{\tabcolsep}{8pt} 
\begin{tabular}{cccc}
\hline
\hline
& $\text{E}(\text{S}_1)$ (eV) & $\text{E}_g$ (eV) & $\text{E}_{\text{binding}}$ (eV) \\
\hline
model blue & $1.88$ & $3.97$ & $2.11$\\
model red & $1.91$ & $4.62$ & $2.71$\\
real red & $2.34$ & $2.86$ & $0.52$\\
\hline
\hline
\end{tabular}
\caption{Values for the first exciton energy $\text{E}(\text{S}_1)$, energy of the fundamental gap $\text{E}_g$ and exciton binding energy $\text{E}_{\text{binding}}$ of the three PDA chains.}
\label{table:binding_energy}
\end{table}

\begin{table}[t]
\centering
  \setlength{\tabcolsep}{8pt} 
\begin{tabular}{ccc}
\hline
\hline
structure & $\Delta$ (eV) - theory &  $\Delta$ (eV) - experiment \\
\hline
blue & $2.23$ (model system) &  $2.64$ \\
red  & $1.06$ & $1.16$ \\
model red & $1.45$ & --- \\
\hline
\hline
\end{tabular}
\caption{Values for the stop-band width of the three disorder-free PDA chains.}
\label{table:Rabi_static}
\end{table}

\section{Static disorder: the effect of finite chain length}
\label{static-disoder}

In a realistic film of PDA, the system deviates from the perfectly
periodic behavior in more than one ways. As we saw when discussing
the experimental reflectivity spectra, unpolymerized chains of diacetylene
can be present, as well as minority populations of chains at different
conformations. Imperfections in the chain polymerization can lead to breaks
in the conjugation of the system, or even to the termination of the polymer
structure after a finite number of monomers. Here we consider such effects
limiting the chain conjugation, and we ignore atomic defects and vacancies,
due to the pure character of the samples as discussed previously. 

\begin{figure}[tb]
\centering
\includegraphics[width=0.4\linewidth]{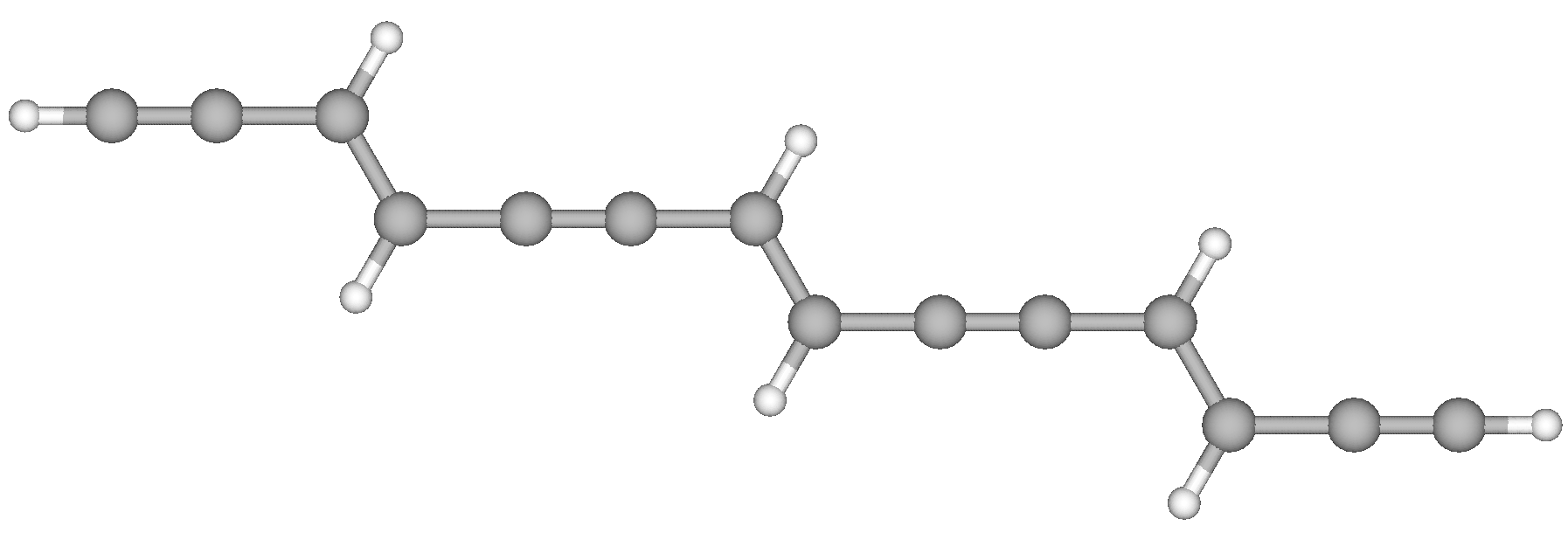}
\caption{A hydrogen-terminated trimer of diacetylene.}
\label{fig:trimer}
\end{figure}

In order to understand the impact of finite conjugation lengths on the exciton and 
polariton properties, we study five different finite systems consisting of 
one to five diacetylene monomers, which we terminate on both sides with 
hydrogen atoms. For simplicity, we choose to terminate the chains after the 
occurrence of a triple bond, so that the carbon atoms at the boundaries
only need to bond with a single hydrogen. As an example, 
Figure\,\ref{fig:trimer} visualizes a trimer of diacetylene. 

\begin{figure}[tb]
\centering
\includegraphics[width=0.4\linewidth]{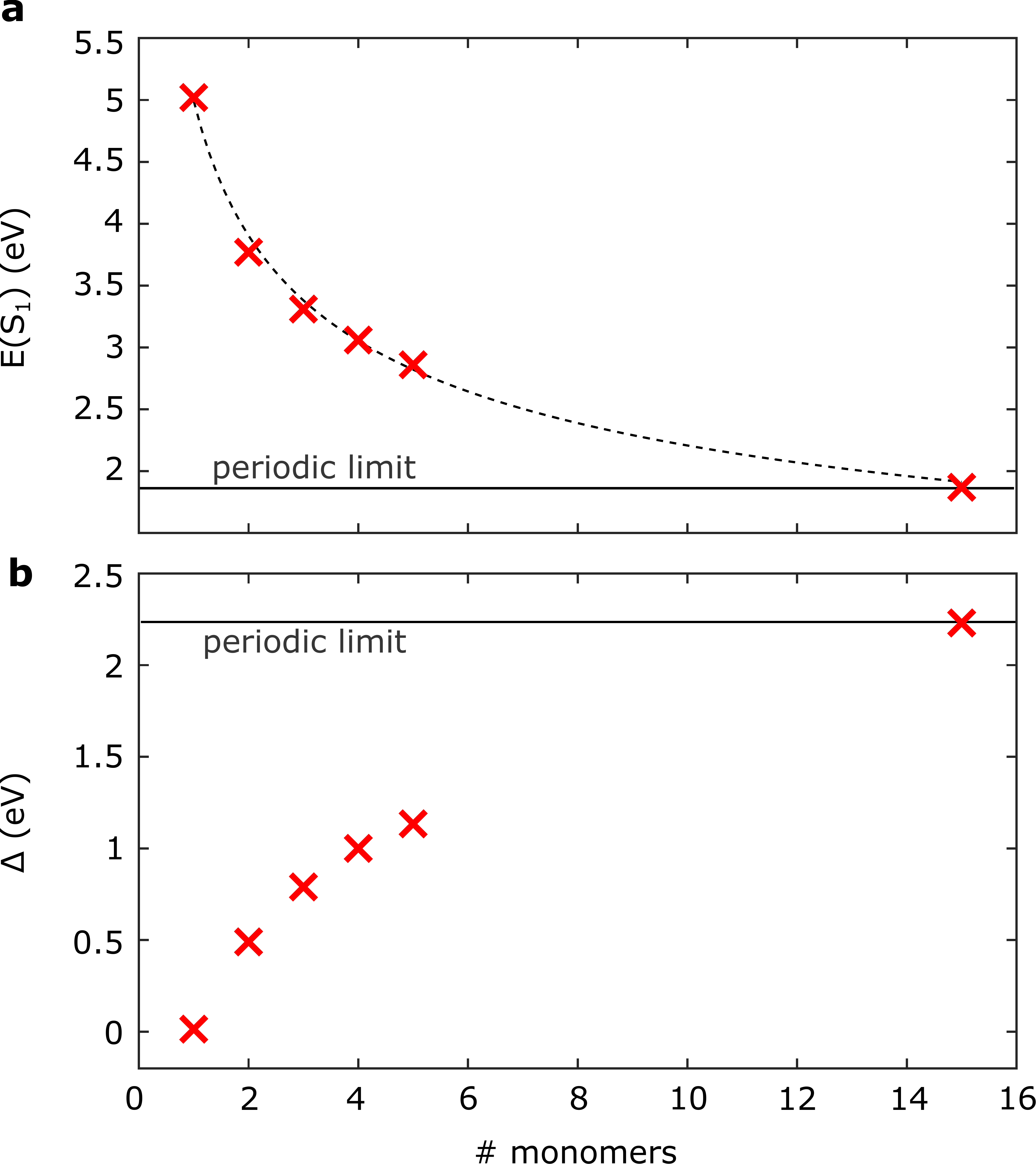}
\caption{The exciton energy (panel \textbf{a}) and stop-band width (panel \textbf{b}) for chains consisting of a different number of diacetylene monomers $N$. Since the constructed oligomers are planar, the periodic limit
of the model blue chains is approached for a larger number of monomers.}
\label{fig:DA_monomers}
\end{figure}

We perform
$GW$-BSE calculations on these oligomers, introducing a vacuum of $20$\,\AA 
\hspace{0.05cm} between their periodic images. We thus obtain $\epsilon(\omega)$ for each
of these, and similar to the previous section, compute the energy of the
first singlet exciton $\text{E}(\text{S}_1)$ and the stop-band width $\Delta$, which we visualize in Figures\,\ref{fig:DA_monomers}a and \,\ref{fig:DA_monomers}b respectively. We have chosen to construct the
oligomer chains so that they are planar, and hence by increasing the number
of monomers, we approach the periodic limit of the model blue PDA chains of
the previous section, which is also noted on Figure\,\ref{fig:DA_monomers}.

First of all, it is clear from Figure\,\ref{fig:DA_monomers}a that as we add
more monomers to the chain, the exciton energy is stabilized and converges
to the periodic limit. Due to the large computational cost, we were unable
to add the required number $N$ of diacetylene monomers in order to reach the
periodic limit, however by fitting a model of the form $\text{E}(\text{S}_1)=a\cdot 
N^b$ to the first five datapoints of Figure\,\ref{fig:DA_monomers}a, we estimate that
the periodic limit is obtained for $N=15$ monomers, in very close agreement with
a previous experimental study \cite{Wudl1986}.
Furthermore, we find that the stop-band width increases with a larger number
of monomers in Figure\,\ref{fig:DA_monomers}b, and the periodic limit
of an infinite chain provides the maximal possible magnitude for 
light-matter interactions. This can be rationalized from the fact that the
exciton frequency appearing in the oscillator strength condition\,\ref{eq:condition}
decreases with increasing number of diacetylene monomers.

We therefore conclude that the presence of static disorder, in this case in 
the form of finite chain lengths, reduces the magnitude of light-matter interactions. 
However, the large measured stop-band width of $\Delta=2.64$\,eV for the blue chains
suggests that conjugation-limiting effects are not important in PDA. Naturally,
breaks in the conjugation and imperfect polymerization do occur, however the results
suggest that the vast majority of chains extend beyond $15$ diacetylene monomers, for
which number the periodic limit is reached, and the maximal light-matter coupling is
obtained. Here we do not compute the effect of static disorder on the reflectance of
PDA, as this would require explicit inclusion of e.g. unpolymerized diacetylene monomers
within a supercell of PDA, making the cost of these calculations prohibitive. We now
proceed to discuss the effects of dynamic disorder on light-matter coupling.

\section{Dynamic disorder: the effect of vibrations}
\label{dynamic-disorder}

A purely electronic structure calculation predicts that there is only
a single bright exciton state in PDA, the vertical transition peak 
in the computed absorption spectrum of Figure\,\ref{fig:PDA_static_absorption}. 
In reality the ground- and excited-state potentials of the PDA chains are displaced along coupled vibrational normal modes, and electronic excitations involve multiple ending vibrational levels (and a thermalized distribution of starting vibrational levels). This
is evident in the experimental absorption spectra of 
Figure\,\ref{fig:PDA_structures},
where one can resolve vibrationally dressed $0-1$ and $0-2$ phonon side bands on top of the purely $0-0$ electronic transition. Intuitively,
it is obvious that these phonon side bands in the structure of 
$\text{Im}(\epsilon(\omega))$, will give rise to more structure in 
$\text{Re}(\epsilon(\omega))$ compared to the case of a single exciton peak,
according to the Kramers-Kronig relation:

\begin{equation}
\label{eq:KK}
\text{Re}(\epsilon(\omega))=P\int_0^{+\infty}\frac{d\nu}{\pi}\text{Im}(\epsilon(\nu))\frac{2\nu}{\nu^2-\omega^2}
\end{equation}

\begin{figure}[tb]
\centering
\includegraphics[width=0.4\linewidth]{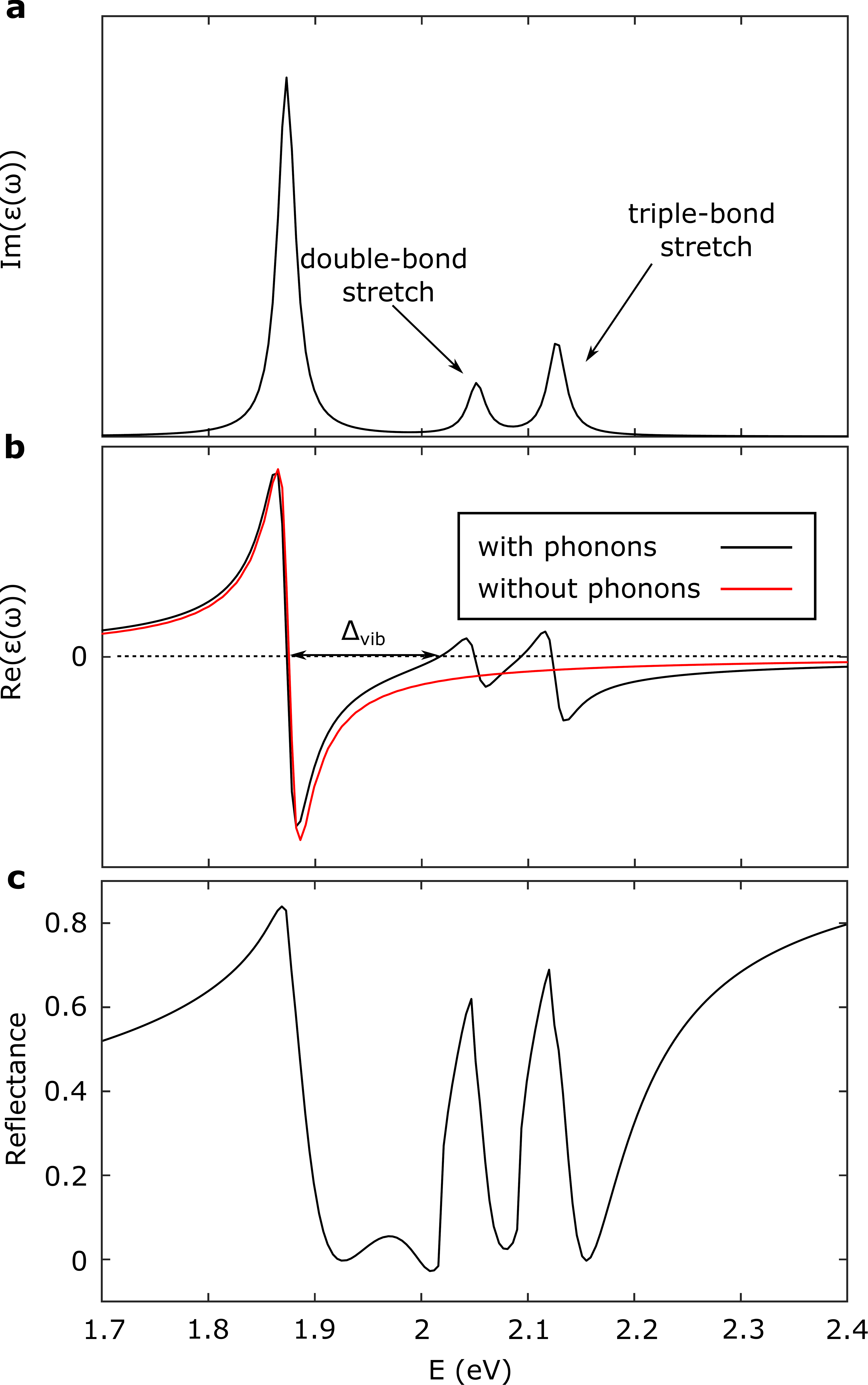}
\caption{The imaginary (panel \textbf{a}) and real part (panel \textbf{b})
of the dielectric response of the model blue chains, once the effect of
phonons is accounted for. The main extra contributions to the structure of $\epsilon(\omega)$, compared to the disorder-free case, arise due to the
effect of the carbon-carbon double- and triple-bond stretch. Comparing the structure
of $\text{Re}(\epsilon(\omega))$ for the cases with and without phonons, we
find that dynamic disorder leads to `breaks' within the region of
$\text{Re}(\epsilon(\omega))<0$. This leads to a richer structure for the reflectance in panel \textbf{c}, compared to the disorder-free case of Figure\,\ref{fig:polaritons_static}.}
\label{fig:vib_progression}
\end{figure}

\noindent
This effect reduces the width of the region in which $\text{Re}(\epsilon(\omega))<0$ following an excitonic transition, as we show in Figure\,\ref{fig:vib_progression}b. We call the width
of this region $\Delta_{\text{vib}}$, and the fact that it is smaller than $\Delta$ means that
a gap in the high reflectivity of the stop-band is introduced (see Figure\,\ref{fig:vib_progression}c below). We therefore proceed to study the effects of
exciton-phonon coupling, as a means of understanding the internal structure of the stop-band.
To this end, we employ the linear-vibronic Hamiltonian \cite{Spano2002}:

\begin{equation}
\label{eq:lin_vib}
H=\text{E}(\text{S}_{\text{1}})\cdot I +\sum_k \hbar \omega_k\lambda_k(b_k^{\dagger}+b_k)+\sum_k\hbar \omega_kb_k^{\dagger}b_k,
\end{equation}

\noindent
where $I$ is the identity, $k$ runs over the phonon modes of the system,
which have frequencies of $\omega_k$. The energy of the first singlet exciton refers to the 
infinite periodic limit. The Huang-Rhys factors are
calculated as:

\begin{equation}
\label{eq:HR_factor}
    \lambda_k=\frac{\text{V}_{\text{ep},k}}{\hbar \omega_k},
\end{equation}

\noindent
where we approximate the first derivative $\text{V}_{\text{ep},k}$ of
the exciton energy along mode $k$ using the finite difference formula:

\begin{equation}
\label{eq:finite_diff}
\text{V}_{\text{ep},k}=\frac{\text{E}(\text{S}_{\text{1}})_{+\delta u(k)}-\text{E}(\text{S}_{\text{1}})_{-\delta u(k)}}{2\delta u(k)},
\end{equation}

\noindent
with $\delta u(k)$ being a small dimensionless displacement along mode $k$.
At a practical level, the calculation of $\text{V}_{\text{ep},k}$ requires
two $GW$-BSE calculations at the displaced configurations of each mode, 
and constitutes a very computationally intensive process. We were therefore
only able to obtain a full parametrization of the Hamiltonian\,\ref{eq:lin_vib} for the two model chains, which only have a
small number of phonon modes due to the small size of their unit cell. The
existence of the side-chains for the real red chains, for which we
previously calculated the exciton properties, makes the task of computing
exciton-phonon interactions a very formidable one, and we thus limit the
following analysis to the model chains.

When computing the Huang-Rhys factors of PDA, one needs to account for the
fact that the exciton can travel a finite distance of $N$ monomers 
coherently, before being scattered by a phonon. Therefore if $\lambda_{k,\text{unit}}$ is the value obtained within a unit cell exciton
calculation where the phonon mode $k$ has been displaced, then the
actual value of the Huang-Rhys parameter is \cite{Spano2015}:

\begin{equation}
\label{eq:Huang-Rhys}
    \lambda_k=\frac{\lambda_{k,\text{unit}}}{N}.
\end{equation}

We estimate the exciton coherence length $N$ as:

\begin{equation}
\label{eq:exc_coh}
    N=\frac{v_{\text{g,exc}}\cdot \tau_{\text{scatter}}}{L_{\text{DA}}},
\end{equation}

\noindent
where $v_{\text{g,exc}}$ the exciton group velocity from
section\,\ref{disorder-free}, $\tau_{\text{scatter}}$ the timescale for the exciton to be
scattered by a phonon, and $L_{\text{DA}}$ the length of a single
diacetylene monomer. Since we find that the excited state surface of
the double-bond stretch is the steepest, i.e. has the largest value of
$\text{V}_{\text{ep},k}$, we approximate $\tau_{\text{scatter}}$ by the
period of $23$\,fs of this mode, giving us an exciton coherence length of $N=15$ monomers for both model chains, in good agreement with previous
estimates on similar systems \cite{Yamagata2011}.

We write the $i^{\text{th}}$ eigenstate of the Hamiltonian of equation\,\ref{eq:lin_vib} as:

\begin{equation}
\label{eq:eigenstates}
    \ket{\Psi^{(i)}} = \sum_{k=1}^M\sum_{\Tilde{v}}^{N_q}c^{(i)}_{\Tilde{v}_k}\ket{1,\Tilde{v}_k}
\end{equation}

\noindent
where there are $M$ vibrational modes in total. The basis state $\ket{1,\Tilde{v}_k}$ is the short-hand notation for

\begin{equation}
    \ket{1,\Tilde{v}_k}=\ket{1,0,0,...,\Tilde{v}_k,0,...,0},
\end{equation}

\noindent
i.e. the state where the exciton is populated and the first index set to 
one, and where among all vibrational modes, only mode $k$ is populated by $\Tilde{v}_k$ quanta. We thus only allow for a single mode
to be excited at one time, as a means of simplifying the problem.
In principle,
every phonon mode can admit an infinite number of quanta, however in
practice we need to truncate the associated Hilbert space to only allow for
up to $N_q$ quanta in each mode. We find that for the model PDA chains 
$N_q=3$ converges the results for the dielectric response. 

Similar to previous works \cite{Spano2002}, we include the effect of vibrations in the
absorption spectrum by taking:

\begin{align}
    \text{Im}(\epsilon(\omega))=(c^{(1)}_0)^2\cdot\text{Im}(\epsilon(\omega))_{0-0}+\nonumber\\ |\sum_{k}\sum_{\Tilde{v}}c^{(1)}_{\Tilde{v}_k}\frac{\lambda_k^{2\Tilde{v_k}}e^{-\lambda_k^2}}{\Tilde{v_k}!}|^2W(\omega-\omega_{\text{S}_1}-\omega_k),
\end{align}

\noindent
where $W(\omega-\omega_{\text{S}_1}-\omega_k)$ a Lorentzian centered around 
$\omega_{\text{S}_1}+\omega_k$, for which we use the same width as for the 
peaks of the purely electronic part $\text{Im}(\epsilon(\omega))_{0-0}$
of the dielectric response, as obtained from $GW$-BSE calculations detailed 
in Appendix B. Moreover, in Appendix B, we show
for the example case of the model blue chains, that the result for the width $\Delta_{\text{vib}}$
is largely independent of
the peak width $\gamma$, for values of $\gamma$ between $5$\,meV and $50$\,meV.
The coefficients $c^{(1)}$ are obtained by
diagonalizing the Hamiltonian\,\ref{eq:lin_vib}, with $c^{(1)}_0$ denoting
the contribution from the purely excitonic basis state $\ket{1,0,0,...,0}$.
Here we reside to approximating the absorbance $\alpha(\omega)$ with 
$\text{Im}(\epsilon(\omega))$. This is a reasonable approximation for determining
$\Delta_{\text{vib}}$, since this quantity depends on the intensity ratio
of the $0-0$, $0-1$, $0-2$ peaks etc. The absorbance $\alpha(\omega)$ and imaginary
dielectric response
$\text{Im}(\epsilon(\omega))$ are proportional to each other, which suggests that
the error we introduce is small.

For the example case of the model blue chains, we visualize the result of including
phonon effects on $\text{Im}(\epsilon(\omega))$ in Figure\,\ref{fig:vib_progression}a, while Figure\,\ref{fig:vib_progression}b
shows the corresponding $\text{Re}(\epsilon(\omega))$ as obtained by using the Kramers-Kronig
relation\,\ref{eq:KK}. For $\text{Re}(\epsilon(\omega))$ we also draw a comparison to the case without phonons. The
peaks that feature prominently in the absorption spectrum are due to
the double- and triple-bond carbon stretching motions. Intuitively, one
can understand why these vibrations are most strongly coupled to the
exciton from the fact that they transiently lead to structures with a 
smaller bond-length alternation, i.e. deviating from the Peierls-distorted
structure that gives conjugated polymers such as PDA their semiconducting
character, and bring the material closer to the metallic limit \cite{Kivelson1985}. It is
hence not surprising that these motions have a large effect on the optical 
properties of the studied PDAs. For the computed exciton coherence length of $N=15$, we obtain $\Delta_{\text{vib}}$
values of $137$\,meV and $123$\,meV for the model blue and the model red
chains respectively. We compute the associated reflectivity spectra in Figure\,\ref{fig:vib_progression}c, where we find that vibronic peaks introduce `gaps'
within the high reflectance region of the stop-band. 

\begin{figure}[tb]
\centering
\includegraphics[width=0.4\linewidth]{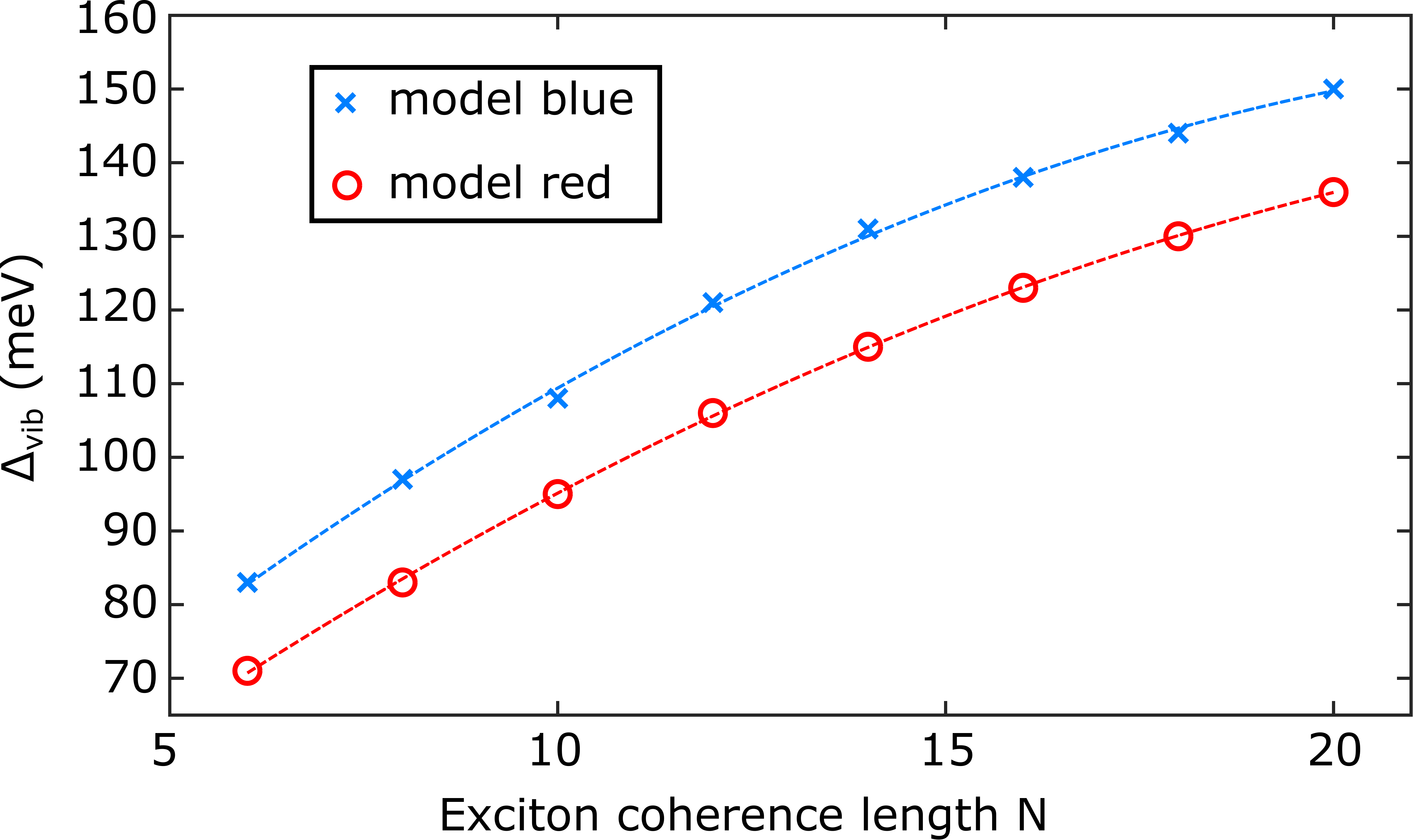}
\caption{The dependence of the width $\Delta_{\text{vib}}$ of the two model PDA chains, on the exciton coherence length.}
\label{fig:exc_coherence}
\end{figure}

Therefore, accounting for vibrational effects allows one to predict part of the internal structure
of the stop-band of an exciton that is strongly coupled to a photon. However, the precise
form of the reflectivity, depends very sensitively on the exciton coherence length $N$, as
shown in Figure\,\ref{fig:exc_coherence}, where we use the width $\Delta_{\text{vib}}$ as
a proxy for the size of phonon-induced gaps in the reflectivity. This
dependence of $\Delta_{\text{vib}}$ on $N$ enters through the Huang-Rhys factors of
the extended system, see equation\,\ref{eq:Huang-Rhys}. In turn, the exciton 
coherence length depends on the exciton group velocity and the period of phonon
modes, as seen in equation\,\ref{eq:exc_coh}. Hence the interplay of phonons with excitons
is established to be crucial for the interaction of the latter with light.

\section{Conclusions}
\label{conclusions}

In this work, we have presented a first principles approach for obtaining
the properties of exciton-polaritons in organic materials. Our theory 
requires input from a calculation of the material's dielectric response,
and is agnostic to the methodology used to obtain it. By employing $GW$-BSE calculations
for the electronic structure of polydiacetylenes, we obtain their stop-band width,
which characterizes the strength of light-matter interactions. The results are in reasonable
agreement with experiment, and we attribute the differences to our treatment of the materials
as perfectly periodic, neglecting effects of disorder and additional excitons
that weakly couple to light.
We compute the speed of propagation of a 
polariton close to the resonance of light with an exciton, showing that this
is an order of magnitude faster than purely excitonic motion. We find the conjugation
length of polymer chains to be a critical parameter towards determining the
magnitude of light-matter interactions, highlighting one manifestation of
how static disorder can lead to weaker exciton-photon coupling. This effect
is not important for the polydiacetylenes studied here, due to the purity of
these materials. It could however be of relevance for different systems. Furthermore, we discuss the important role of dynamic disorder
and show that vibrational effects introduce gaps within
the frequency range of large reflectance that characterizes polaritons. 
We also discuss the screening of the Coulomb 
interaction provided
by the side-chains of polydiacetylene as another potentially important contribution
towards determining the magnitude of light-matter interactions. 
Thus all of the aforementioned factors appearing in realistic 
materials are important towards an accurate description of exciton-photon
coupling. In principle, one needs to account for the above factors at the same
time and self-consistently, something which is computationally challenging and will be
the subject of a future work.

\section*{Acknowledgments}
The authors 
acknowledge useful discussions with Richard Phillips (Cavendish) and Girish Lakhwani (University of Sydney). We thank Gianni Jacucci and Silvia Vignolini (Cambridge) for assistance with reflectivity measurements. We acknowledge
the support of the Winton Programme for the Physics of 
Sustainability. A.M.A. acknowledges the support of the Engineering and Physical 
Sciences Research Council (EPSRC) for funding under grant EP/L015552/1. A.W.C. acknowledges financial support from Agence Nationale de la Recherche (Grant N. ANR-19-CE24-0028). The work in Mons is supported by the Belgian National Fund for Scientific Research (FRS-FNRS). Computational resources were provided by the Consortium des Équipements de Calcul Intensif (CECI). DB is FNRS Research Director.

\section*{Data availability}

\noindent
The data underlying this publication can be found in [URL added in proof].

\section*{Appendix}
\subsection*{A: Experimental reflectivity measurements}
\label{experimental}

Microscopic transmission and reflectance measurements were performed  using  a  Zeiss AxioScope  optical  microscope  in  K{\"o}hler  illumination  equipped with a 100$\times$ objective  (Zeiss  EC Epiplan-APOCHROMAT  0.95  HD  DIC)  coupled to  a  spectrometer  (Avantes  HS2048)  via  an optical  fibre  (Thorlabs,  FC-UV50-2-SR). Five spectra  were  collected  for  each  sample  using an integration time of 10 ms and 20 ms for reflection and transmission measurements, respectively. The reflectance and transmittance were calculated using a silver mirror (ThorLabs, PF10-03-P01) and the glass substrate as references respectively.

\subsection*{B: Computational details}
\label{computational}

\subsection*{Electronic structure calculations}

We perform DFT calculations using the Quantum Espresso software package 
\cite{Giannozzi2009}, with the PBE functional together with a $60$\,Ry plane wave cutoff
energy. For the model blue and model red chains, we employ $8\times4\times2$ and 
$4\times4\times2$ $\mathbf{k}$-point grids respectively, while for the real red chains, 
we use $4\times4\times2$. For the model PDAs, we introduce a vacuum of $10$\,\AA 
\hspace{0.05cm} between neighboring polymer chains, in order to minimize inter-chain
interactions.

We perform energy self-consistent $GW$ calculations (i.e. $GW_0$) using the {\sc yambo} 
code \cite{Marini2009}. For the real chains, we find that including $400$ Kohn-Sham states, $300$ bands for 
the calculation of the polarization function, and a $7$\,Ry cutoff for the dielectric 
matrix, leads to converged values for the quasiparticle bandgaps. For the model blue
(red) chains, we use $40$ ($80$) Kohn-Sham states, $40$ ($80$) bands for 
the calculation of the polarization function, and a $3$\,Ry cutoff for the dielectric 
matrix.

We solve the Bethe-Salpeter equation for the model blue (red) PDA using $4$ ($8$) 
occupied and $4$ ($8$) 
unoccupied bands, converging the position of the first exciton peak. For the exchange 
term in the Bethe-Salpeter kernel, the cutoff is set to $60$\,Ry. For the real red
chains, we use $10$ occupied and $10$ unoccupied bands, and set the exchange to 
$40$\,Ry. In order to better
emulate the experimental conditions, the incident electric field is averaged among the
three spatial directions. The imaginary part of $\epsilon(\omega)$ is obtained by 
fitting a Lorentzian of $10$\,meV width to the computed excitonic eigenvalues of the
BSE Hamiltonian. The broadening of the absorption spectrum due to vibrational effects
as presented in Figure\,\ref{fig:vib_progression} is then added on top of that, along
with the additional peaks. 

Due to the one-dimensional character of the model PDAs, divergences can occur at small 
values of $q$ (i.e. at long distances) in the Coulomb term that appears in the many-body
calculations. In order to avoid this problem, we use the random integration method, as
implemented in {\sc yambo}. We converge the calculations using $10^5$ random q-points in
the first Brillouin zone, and by setting the cutoff for the real space components
of the Coulomb interaction to $20$\,Ry.

For the model PDAs, we estimate the dispersion of the exciton band along 
$\Gamma-X$ as follows. We generate a supercell of size $2\times1\times1$, thus
halving the size of the first Brillouin zone. As a result, the zone-boundary $X$ point
folds onto $\Gamma$, which appears as an extra exciton state once the BSE Hamiltonian
of the $2\times1\times1$ system is diagonalized. The distance of this new band from
the exciton value at $\Gamma$ gives the width of the exciton band, which we find to be
equal to $1.33$\,eV. 

All the aforementioned electronic structure calculations are performed on the PDA
structures as obtained from X-ray diffraction (for the real red system), or
on the structures as obtained from X-ray diffraction and following the substitution
of the side-chains with hydrogen atoms. A geometry optimization of these structures
is not carried out, as the bond-length alternation of these one-dimensional carbon
chains is known to be extremely sensitive to the fraction of exact exchange that
the employed DFT functional includes\cite{Mostaani2016}. It has been shown that
while hybrid functionals perform better than GGA\cite{Mostaani2016}, it is necessary
to use very accurate methods such as Quantum Monte Carlo in order to obtain
geometries with realistic bond-length alternations, which is beyond the scope of this
work. As discussed in the main manuscript, the exciton energy is extremely sensitive
to the precise value of the bond-length alternation. Therefore we limit ourselves
to obtaining exciton energies for unoptimized structures, and only optimize
the geometry of the studied structures prior to phonon calculations, 

\subsection*{Phonon calculations}

We calculate the lattice dynamics of the different PDA variants in the 
harmonic approximation using finite differences \cite{Kunc1982,Monserrat2018}. We calculate the 
dynamical matrix on a $1\times1\times1$ coarse $\mathbf{q}$-point grid. Therefore, we
only account for the effect of $\Gamma$-point phonons, which provides a good 
approximation, since these are most strongly coupled to the zero-momentum exciton
states that result from photoexcitation. Prior to phonon calculations, the geometry
of the different PDAs is optimized using the PBE functional. 

\subsection*{Dependence of polariton splitting on peak width}

\begin{figure}[tb]
\centering
\includegraphics[width=0.4\linewidth]{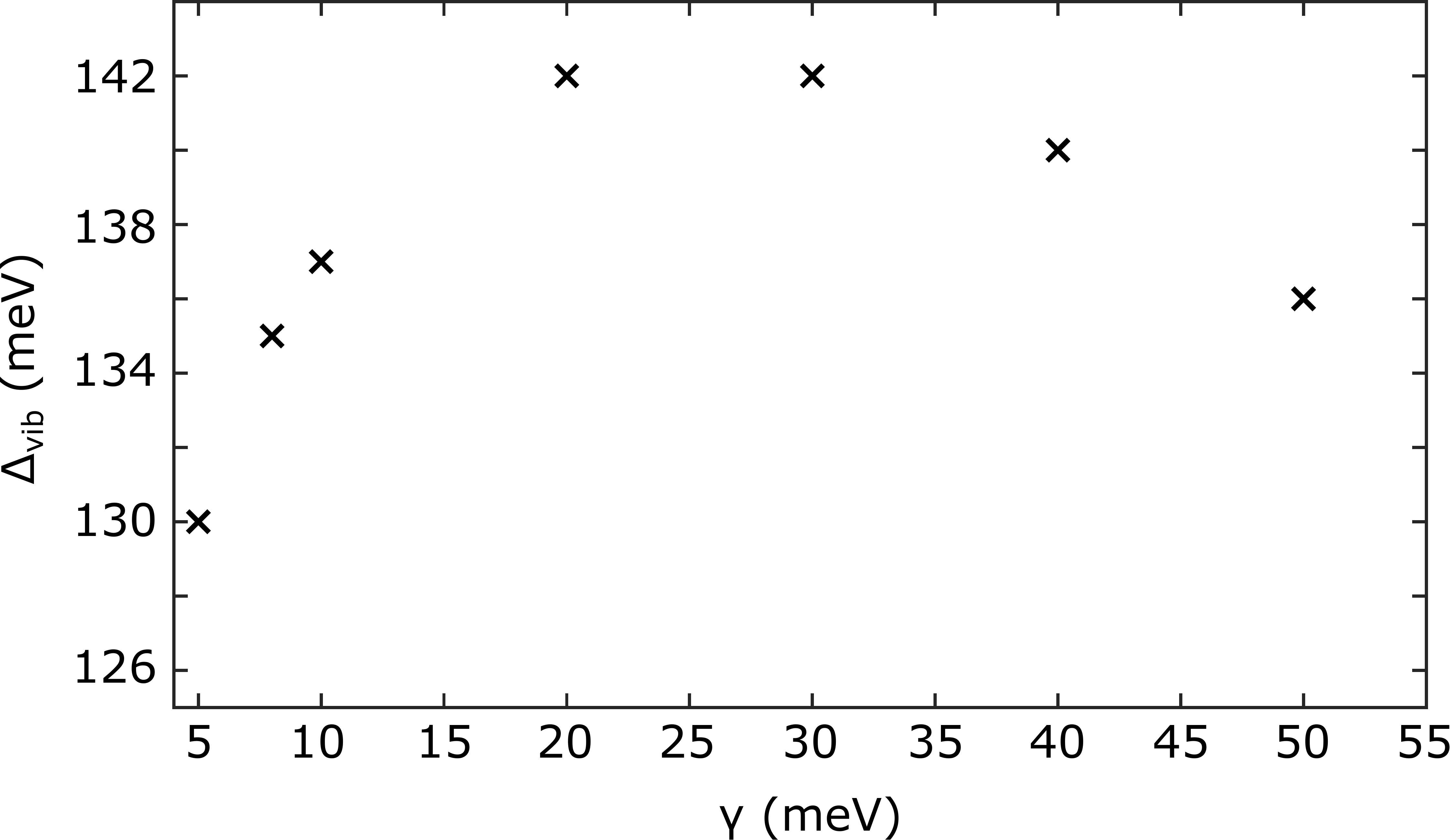}
\caption{The dependence of the width $\Delta_{\text{vib}}$ of the first region with $\text{Re}(\epsilon(\omega))$ on the width $\gamma$ of the exciton
and vibronic peaks.}
\label{fig:Rabi_gamma}
\end{figure}

Figure\,\ref{fig:Rabi_gamma} visualizes the value of $\Delta_{\text{vib}}$ of the model blue PDA,
within a range of values for the width of the exciton and vibronic peaks, ranging from 
$5$\,meV to $50$\,meV. We find that $\Delta_{\text{vib}}$ does not vary by more than $10$\,meV
within this range of widths. Here an exciton coherence length $N=15$ has been used.

\textbf{}
\bibliographystyle{unsrt}
\bibliography{references}

\end{document}